\documentclass[aps,pre,twocolumn,groupedaddress]{revtex4-1}
\usepackage{graphicx}
\usepackage{dcolumn}
\usepackage{bm}
\usepackage{amsmath}
\usepackage{amssymb}
\usepackage{color}

\begin{document}


\title{Detachment of fluid membrane from substrate and vesiculations}

\author{Hiroshi Noguchi${^{a,b}}$}
\email[]{noguchi@issp.u-tokyo.ac.jp}
\affiliation{
${^a}$ Institute for Solid State Physics, University of Tokyo,
 Kashiwa, Chiba 277-8581, Japan.
${^b}$ Institut Lumi{\`e}re Mati{\`e}re, UMR5306 Universit{\'e} Lyon 1-CNRS, Universit{\'e} de Lyon 69622 Villeurbanne, France.}

\date{\today}

\begin{abstract}
The detachment dynamics of a fluid membrane with an isotropic spontaneous curvature 
from a flat substrate are studied by using meshless membrane simulations.
The membrane is detached from an open edge leading to vesicle formation.
With strong adhesion, the competition between the bending and adhesion energies
determines the minimum value of the spontaneous curvature  for the detachment.
In contrast, with weak adhesion,
a detachment occurs at smaller spontaneous curvatures due to the membrane thermal undulation.
When parts of the membrane are pinned on the substrate,
the detachment becomes slower and
a remained membrane patch forms straight or concave membrane edges.
The edge undulation induces vesiculation of long strips and disk-shaped patches. 
Therefore, membrane rolling is obtained only for membrane strips shorter than the wavelength for deformation into unduloid.
This suggests that the rolling observed for Ca$^{2+}$-dependent membrane-binding proteins, annexins A3, A4, A5, and A13,
results from by the anisotropic spontaneous curvature induced by the proteins.
\end{abstract}

\maketitle

\section{Introduction}

Lipid membranes supported on a solid substrate
 are considered as a model system for biological membranes
and are extensively used 
to study immune reaction and protein functions 
as well as membrane properties \cite{tana05,cast06,acha10,ales14,weer15}.
Membranes are placed on a solid or polymer layer, and
 a wide range of the surface-specific analytical techniques can be applied.

The adhesion of vesicles onto a substrate is a typical method for producing supported membranes.
For the establishment of this method, the adhesion process has been investigated intensively
by experiments~\cite{john02,cha06,hain13,mapa18} and 
coarse-grained molecular simulations~\cite{fuhr13,kong16,wen17}.
A vesicle adheres to the substrate, and the resultant high surface tension induces membrane rupture, 
leading to membrane spreading on the substrate. 
Moreover, the adhesion of bicelles can result in the formation of supported membranes~\cite{kola17,yama18}.

The opposite process, namely detachment, has been the subject of few studies.
However, Boye {\it et al.} very recently reported 
that the annexin proteins can detach lipid membranes from a substrate~\cite{boye17,boye18}. 
Various types of detachment dynamics were observed.
The annexins are Ca$^{2+}$-dependent membrane-binding proteins and
have functions in endo/exocytosis and membrane repair~\cite{gerk05,bout15,blaz15}.
The annexins A3, A4, A5, and A13 induce  membrane rolling from open edges.
In particular, thick rolls are grown with A4 and A5, while thin branched rolls are formed
with A3 and A13. 
The annexins A1 and A2 induce membrane blebbing and folding.
The bleb exhibits a spherical shape and remains still connected to the membrane patch.
Moreover, the relation between the observed rolling and the membrane repair
has been discussed in Ref.~\citenum{boye17}.

Herein, the detachment dynamics of membrane patches from a flat substrate is studied by
using meshless membrane simulations.
Since the  annexins induce the rolling and blebbing,
they almost certainly bends the bound membrane.
The folding indicates that the  annexins A1 and A2 also bind two membranes.
Furthermore, the protein--protein interaction is also important.
When the trimer formation of the annexins A4 is inhibited, 
the membrane rolling does not occur~\cite{boye17}.
Annexins A5 forms a two-dimensional ordered array on the membrane~\cite{bout15,bout11},
so that it may effectively solidify the membrane.
Thus, the annexins definitely play various roles in the membrane detachment.
However, in this study, the isotropic spontaneous curvature is considered as the minimum role,
since the membrane bending is required for the detachment.
Certain proteins, such as the Bin/Amphiphysin/Rvs (BAR) superfamily proteins,
are known to bend along the domain; that is, induce anisotropic spontaneous curvature~\cite{mcma05,suet14,simu15a,joha15}.
However, to our knowledge,  experimental evidence of the annexins bending the membrane anisotropically
 has not yet been reported.
Herein, the different types of detachment dynamics occurring on a membrane with the isotropic spontaneous curvatures
are studied by means of simulation.
The isotropic spontaneous curvature can also be induced by the polymer anchoring and colloid adhesion~\cite{phil09,lipo13,dasg17}.
Moreover, the pinning effects on the membrane detachment are investigated.
In experiments, parts of the membrane patch are often pinned in the original position.
Finally, the additional requirements for obtaining different types of dynamics are discussed.

The simulation model and method are described in Sec.~\ref{sec:method}.
Various types of membrane models have been developed for membrane simulations~\cite{muel06,vent06,nogu09}.
In this study, a spin type of the meshless membrane models 
is used, in which membrane particles self-assemble into a membrane~\cite{shib11}.
This model was previously applied to membranes with an isotropic spontaneous curvature~\cite{nogu16a,nogu19}
or anisotropic spontaneous curvature~\cite{nogu14,nogu16,nogu17,nogu19a}, as well as their mixture~\cite{nogu17a}.
One can efficiently simulate membrane deformation with membrane fusion and fission
in a wide range of membrane elastic parameters.
The simulation results of the membrane detachment without and with pinning
are described in Secs.~\ref{sec:detach} and \ref{sec:pin}, respectively.
Finally, Sec.~\ref{sec:sum} presents a summary and discussion of this study.

\section{Simulation model and method}~\label{sec:method}

A fluid membrane is represented by a self-assembled one-layer sheet of $N$ particles.
The position and orientational vectors of the $i$-th particle are ${\bf r}_{i}$ and ${\bf u}_i$, respectively.
Since the  details of the meshless membrane model are described at length in Ref.~\citenum{shib11},
 it is only briefly presented here.

The membrane particles interact with each other via the potential $U=U_{\rm {rep}}+U_{\rm {att}}+U_{\rm {bend}}+U_{\rm {tilt}}$.
The potential $U_{\rm {rep}}$ is an excluded volume interaction with a diameter $\sigma$ for all particle pairs.
The solvent is implicitly accounted for by an effective attractive potential as follows:
\begin{eqnarray} \label{eq:U_att}
\frac{U_{\rm {att}}}{k_{\rm B}T} =  \frac{\varepsilon_{\rm mb}}{4}\sum_{i} \ln[1+\exp\{-4(\rho_i-\rho^*)\}]- C,
\end{eqnarray} 
where  $\rho_i= \sum_{j \ne i} f_{\rm {cut}}(r_{i,j})$, $C$ is a constant,
 $k_{\rm B}T$ is the thermal energy, and $\rho^*$ is the characteristic density with $\rho^*=6$.
$f_{\rm {cut}}(r)$ is a $C^{\infty}$ cutoff function
 and $r_{i,j}=|{\bf r}_{i,j}|$ with ${\bf r}_{i,j}={\bf r}_{i}-{\bf r}_j$:
\begin{equation} \label{eq:cutoff}
f_{\rm {cut}}(r)=\left\{ 
\begin{array}{ll}
\exp\{A(1+\frac{1}{(r/r_{\rm {cut}})^n -1})\}
& (r < r_{\rm {cut}}) \\
0  & (r \ge r_{\rm {cut}}) 
\end{array}
\right.
\end{equation}
where $n=12$, $A=\ln(2) \{(r_{\rm {cut}}/r_{\rm {att}})^n-1\}$,
$r_{\rm {att}}= 1.8\sigma$, and $r_{\rm {cut}}=2.1\sigma$.
For low densities,
$U_{\rm {att}}$ is a pairwise attractive potential, while the attraction is
smoothly truncated at $\rho_i \gtrsim \rho^*$.
The bending and tilt potentials
are given by 
\begin{eqnarray} \label{eq:bend}
\frac{U_{\rm {bend}}}{k_{\rm B}T} &=& \frac{k_{\rm {bend}}}{2} \sum_{i<j} ({\bf u}_{i} - {\bf u}_{j} - C_{\rm {bd}} \hat{\bf r}_{i,j} )^2 w_{\rm {cv}}(r_{i,j}), \\
\frac{U_{\rm {tilt}}}{k_{\rm B}T} &=& \frac{k_{\rm{tilt}}}{2} \sum_{i<j} [ ( {\bf u}_{i}\cdot \hat{\bf r}_{i,j})^2
 + ({\bf u}_{j}\cdot \hat{\bf r}_{i,j})^2  ] w_{\rm {cv}}(r_{i,j}), 
\end{eqnarray}
where
 $\hat{\bf r}_{i,j}={\bf r}_{i,j}/r_{i,j}$ and $w_{\rm {cv}}(r_{i,j})$ is a weight function.

The membrane elastic properties for a wide range of parameter sets
 are provided in Refs.~\citenum{shib11} and \citenum{nogu19}.
We employ typical values of membrane properties in lipid membranes in this study.
The spontaneous curvature of the membrane is determined by $C_{\rm {bd}}$ in Eq.~(\ref{eq:bend}).
To detach the membrane, negative values of the spontaneous curvature are used 
and the amplitude of the negative spontaneous curvature is given by $C_0\sigma= -C_{\rm {bd}}/2$.
The bending rigidity $\kappa$ is linearly dependent on $k_{\rm {bend}}$ and $k_{\rm{tilt}}$
but independent of  $\varepsilon_{\rm mb}$ for $\kappa/k_{\rm B}T \gtrsim 20$.
On the other hand, the line tension $\Gamma$ of the membrane edge is linearly dependent on $\varepsilon_{\rm mb}$
but  independent of $k_{\rm {bend}}$ and $k_{\rm{tilt}}$.
Therefore, these two quantities are controlled individually.
Here, we fix the ratio as $k_{\rm {bend}}=k_{\rm{tilt}}=k$:
$\kappa/k_{\rm B}T= 16$, $34$, and
$52$ for  $k=10$, $20$, and $30$ at $\varepsilon_{\rm mb}=4$, respectively. 
The edge tension $\Gamma\sigma/k_{\rm B}T= 3.89$, $5.1$, $6.2$, and $7.3$
 for $\varepsilon_{\rm mb}=4$, $5$, $6$, and $7$  at $k=20$, respectively.
Experimentally, the edge tension is estimated from the membrane pore formation: 
$\Gamma = 4$--$40$pN.~\cite{zhel93,kara03,port10}
When the particle diameter is considered as membrane thickness, $\sigma \simeq 5$nm,
$k_{\rm B}T/\sigma \simeq 1$pN. 
Unless otherwise specified,  $\varepsilon_{\rm mb}=4$ is used in this study.
The ratio of the Gaussian modulus $\bar{\kappa}$ to $\kappa$ is constant~\cite{nogu19}: $\bar{\kappa}/\kappa=-0.9\pm 0.1$.
The area $a_0$ per particle in the tensionless membranes is slightly dependent on $C_0$
as follows: $a_0/\sigma^2= A + B(C_0\sigma)^2$ with
$\{A,B\}=\{1.47,1.1\}$, $\{1.5,2.5\}$, and $\{1.5,3.9\}$ for $k=10$, $20$, and $30$ at $\varepsilon_{\rm mb}=4$,
respectively.

\begin{figure}
\includegraphics{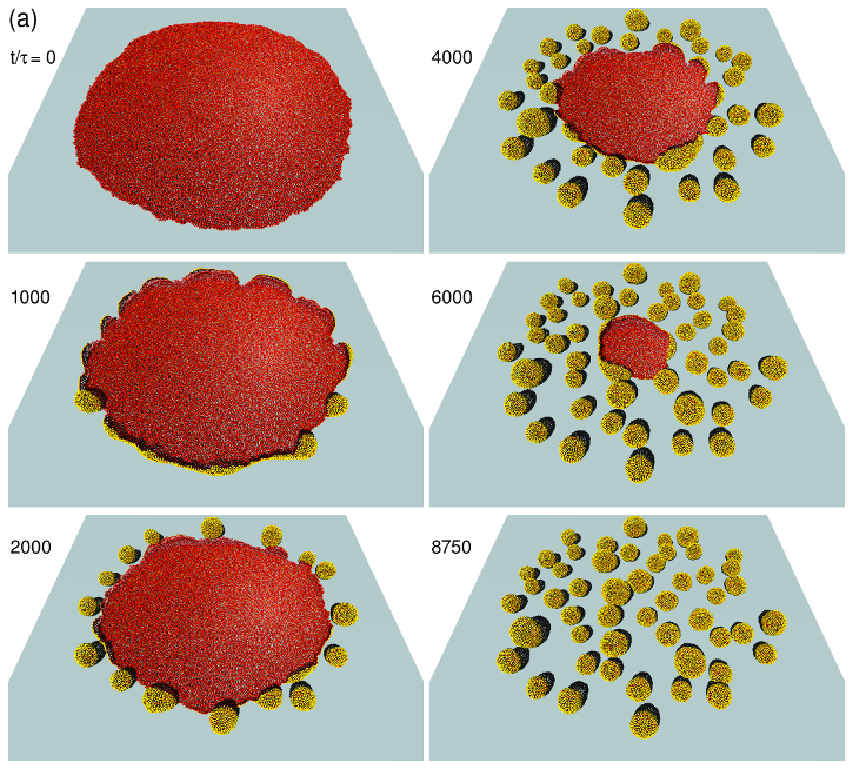}
\includegraphics{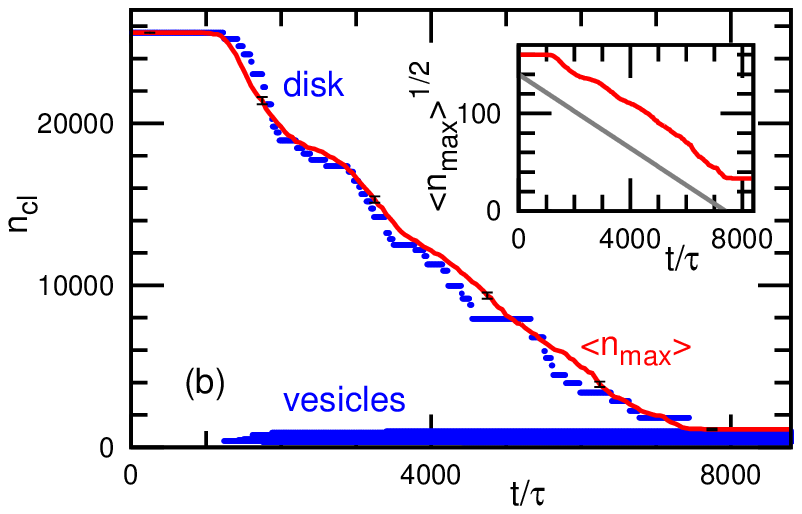}
\caption{
Vesiculation dynamics from the disk-shaped membrane on a solid substrate at
 $C_0\sigma=0.16$, $\kappa/k_{\rm B}T=34$, $\varepsilon_{\rm ad}/k_{\rm B}T=1$, and $N=25~600$.
(a) Sequential snapshots at $t/\tau=0$, $1000$, $2000$, $4000$, $6000$, and $8750$.
The orientation vector ${\bf u}_i$ of a membrane particle lies along the direction 
from the yellow to red hemispheres.
(b) Time evolution of the cluster sizes $n_{\rm cl}$.
The blue dots indicate the data corresponding to the snapshots in (a).
The red lines denote the average of the maximum cluster size $\langle n_{\rm max}\rangle$.
The error bars are displayed at several data points.
In the inset, the red line illustrates $\langle n_{\rm max}\rangle^{1/2}$.
The gray straight line is  guide to the eye.
}
\label{fig:cn}
\end{figure}

The solid substrate is set to $z=0$
and interacts with the membrane particles via the Lennard-Jones potential:
\begin{equation}
U_{\rm ad}=\sum 4\varepsilon_{\rm ad}\Big[\Big(\frac{\sigma}{z_{i}}\Big)^{12}-\Big(\frac{\sigma}{z_{i}}\Big)^6\Big],
\end{equation}
which has an energy minimum at $z=2^{1/6}\sigma$  with a depth of $\varepsilon_{\rm ad}$.
The membrane is initially equilibrated at $C_0=0$.
Further, $C_0$ is transformed to a target value at $t=0$.
Membrane disks are simulated for $N=6400$ and $25~600$
with radii of $R_{\rm disk}/\sigma= 55$ and $110$, respectively, at $C_0=0$, $k=20$, and $\varepsilon_{\rm mb}=4$.
The limit of a large membrane disk, namely
a membrane strip, is also simulated, in which the membrane is connected to itself by the periodic boundary condition on the $x$ axis.
The strip length is $L_x=160\sigma$ with $N=25~600$ unless otherwise specified.

To study the pinning effects, 
$N_{\rm pin}$ membrane particles are fixed in the initial position
at $k=20$, $\varepsilon_{\rm mb}=6$, and $N=25~600$.
The detachment dynamics are simulated at $N_{\rm pin}=25$, $50$, $100$, and $200$,
that is, the density of the pinned particle, $\phi_{\rm pin} =N_{\rm pin}/N=  0.001$, $0.002$, $0.0039$, and $0.0078$.
The pinned particles are randomly selected for each simulation run.
Since the pinned particles are often separated from the membrane during the detachment
at the smaller attraction of $\varepsilon_{\rm mb}=4$ for $\varepsilon_{\rm ad}/k_{\rm B}T=1$, this attraction strength is selected.

Molecular dynamics with a Langevin thermostat is employed~\cite{shib11,nogu11}.
The numerical errors of the phase boundaries and time evolution are estimated from $3$ and $20$ independent runs, respectively.
In the following sections, the results are displayed with the diameter of membrane particles $\sigma$ as the length unit
 and $\tau= \sigma^2/D_0$ as the time unit,
where $D_0$ is the diffusion coefficient of isolated membrane particles.

\begin{figure}
\includegraphics{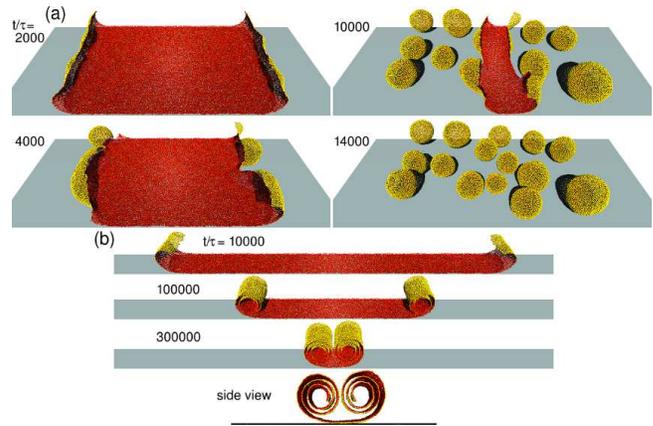}
\caption{
Vesiculation and rolling of membrane strips at
 $C_0\sigma=0.1$, $\kappa/k_{\rm B}T=34$, and $\varepsilon_{\rm ad}/k_{\rm B}T=0.1$.
(a) $L_x/\sigma=160$ and $N=25~600$. (b)  $L_x/\sigma=40$ and $N=12~800$.
The side view is also presented at $t/\tau=300~000$.
}
\label{fig:roll}
\end{figure}

\section{Detachment of Fluid Membrane without Pinning} \label{sec:detach}

First,  the detachment of a homogeneous fluid membrane with a spontaneous curvature is considered.
The membrane has a constant isotropic spontaneous curvature, $-C_0$ 
and all membrane particles are mobile.
When a high curvature $C_0$ is imposed, a membrane disk is detached from the substrate,
and vesicles are formed [see Fig.~\ref{fig:cn}(a) and Movie~S1 provided in ESI].
The membrane edge is rolled up, and the edge undulation expands leading to the vesicle formation.
Under this condition, the average vesicle size is $\langle n_{\rm ves} \rangle=530$,
which is several times larger than the minimum vesicle size formed 
by the membrane closure of a cup-shaped membrane patch.
The membrane sizes of transition and spinodal decomposition 
 are  $\{N_{\rm tra},N_{\rm spi}\}=\{83,110\}$ for the cup-to-vesicle transition~\cite{nogu19}.
During the division into small vesicles, the edge length increases temporally 
[see the middle-left snapshot in Fig.~\ref{fig:cn}(a)]. 
This differs from the cup-to-vesicle transition, in which the edge length decreases monotonically.

The time evolution of the clusters is shown in Fig.~\ref{fig:cn}(b).
When membrane particles are closer than $r_{\rm {att}}$, it is considered that they belong to the same cluster.
Since several vesicles are simultaneously formed along the membrane edge [see the snapshot at $t/\tau=2000$ in Fig.~\ref{fig:cn}(a)],
a stepwise decrease in the size of the maximum cluster (membrane disk) appears.
On average, the maximum cluster size decreases as 
\begin{equation}\label{eq:nmax}
\langle n_{\rm max} \rangle^{1/2} \simeq N^{1/2} - b t
\end{equation}
as shown in the inset of Fig.~\ref{fig:cn}(b).
This is due to the edge length decrease as $L_{\rm edge} \simeq \sqrt{n_{\rm max}}$.
The size approximately decreases as  $dn_{\rm max}/dt= - 2b \sqrt{n_{\rm max}}$
so that Eq.~(\ref{eq:nmax}) is obtained.
A similar decrease was obtained during membrane lysis, 
when the membrane dissolution occurs only from the membrane edge~\cite{nogu06a}.

These vesiculation dynamics are not qualitatively modified for long membrane strips and small membrane disks of $N=6400$.
The membrane strip is simulated to investigate the straight edge as the limit of the large membrane disk.
At $L_x/\sigma \geq 80$, the bent edges begin to undulate and vesicles are formed at $C_0\sigma=0.1$
[see Fig.~\ref{fig:roll}(a) and Movie~S2 provided in ESI].
However, a rolled membrane is formed on a short strip of $L_x/\sigma = 40$
[see Fig.~\ref{fig:roll}(b) and Movie~S3 provided in ESI].
At $L_x/\sigma = 160$, the membrane edge undulates into two or three bumps 
[see the snapshot at $t/\tau=4000$ in Fig.~\ref{fig:roll}(a)],
and typically each bump grows into one vesicle.
This edge undulation is suppressed at $L_x/\sigma = 40$.
Although various conditions are examined, this rolling is obtained only for the short strips.
Thus, the straight free edges are always unstabilized for the longer strips.
This roll unstabilization is caused by the formation of an unduloid-shaped membrane, 
which periodically undulates along the rotational axis
and exhibits a constant mean curvature~\cite{kenm03,nait95}.
The cylinder with a radius of $1/C_0$ can be continuously transformed into unduloids by maintaining the mean curvature,
and the transformation begins at a wavelength of $l_{\rm und}=2\pi/C_0$.
Hence, the unduloid formation is suppressed for strips that are sufficiently shorter than $l_{\rm und}$.
This threshold length agrees with the simulation results, since $l_{\rm und}/\sigma \simeq 60$ at  $C_0\sigma=0.1$.
A similar instability of a tubular lipid vesicle was observed in polymer anchoring~\cite{tsaf01}.

\begin{figure}
\includegraphics{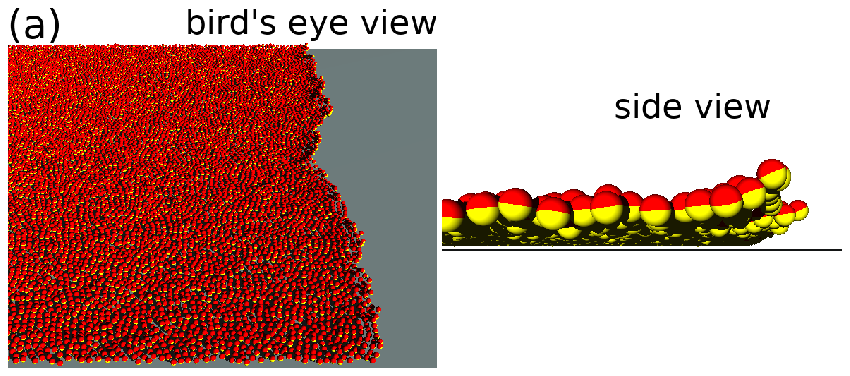}
\includegraphics{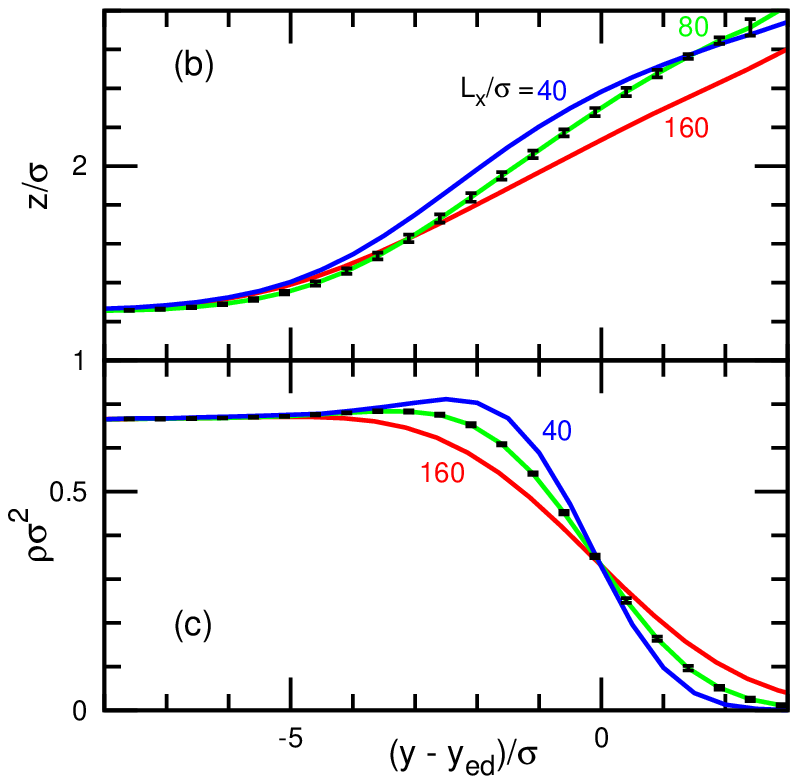}
\caption{
Membrane edge of the undetached membrane strips at  $C_0\sigma=0.1$, $\kappa/k_{\rm B}T=34$, and $\varepsilon_{\rm ad}/k_{\rm B}T=0.5$.
(a) Snapshots from bird's eye and side views  at $L_x/\sigma=160$.
(b) Membrane height $z(y)$ and (c) density $\rho(y)$ at $L_x/\sigma=40$, $80$, and $160$.
The error bars are shown for several data points at $L_x/\sigma=80$.
}
\label{fig:height1}
\end{figure}

\begin{figure}
\includegraphics{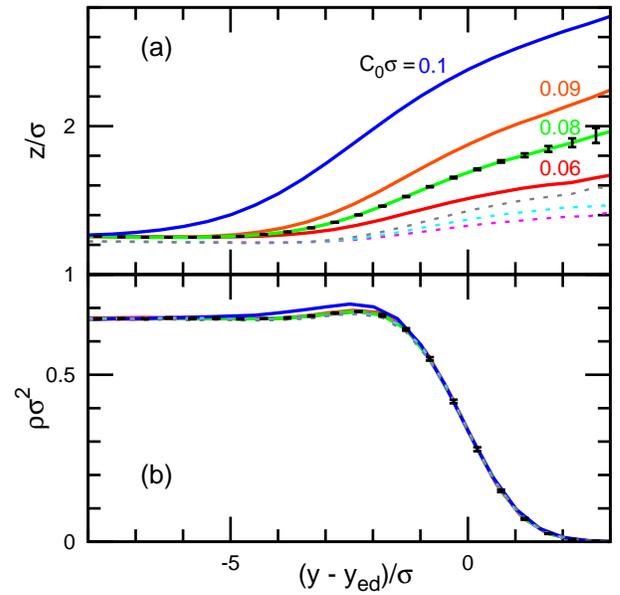}
\caption{
Membrane height $z(y)$ of the undetached membrane strips at $\kappa/k_{\rm B}T=34$ and $L_x/\sigma=40$.
The solid lines indicate the data at $C_0\sigma=0.06$, $0.08$, $0.09$, and $0.1$ for  $\varepsilon_{\rm ad}/k_{\rm B}T=0.5$.
The dashed lines indicate the data at $C_0\sigma=0.06$, $0.08$, and $0.1$ from the bottom to top for $\varepsilon_{\rm ad}/k_{\rm B}T=1$.
The error bars are indicated for several data points at $C_0\sigma=0.08$ and $\varepsilon_{\rm ad}/k_{\rm B}T=0.5$.
}
\label{fig:height2}
\end{figure}

At a small $C_0$, the membrane remains on the substrate.
The edges are slightly separated from the substrate and fluctuate along the edge ($x$ axis) as shown in Fig.~\ref{fig:height1}(a).
To quantitatively evaluate the edge shape, the membrane height profile $z(y)$ perpendicular to the edge is calculated 
and presented in Figs.~\ref{fig:height1}(b) and \ref{fig:height2}(a).
The edge position $y_{\rm ed}$ is defined as at the position 
in which the density $\rho$ is half of that of the middle membrane region.
Since the edge $y$ position fluctuates along the $x$ axis, the changes in the density and height
become more gradual for longer strips [see Figs.~\ref{fig:height1}(b) and (c)].
As $C_0$ increases or $\varepsilon_{\rm ad}$ decreases, the edge exhibits greater bending [see Fig.~\ref{fig:height2}(a)].
It is noted that the height profile at $y\gtrsim y_{\rm ed}$ can be significantly modified by the selection of the averaging methods.
At  $y\gtrsim y_{\rm ed}$, the slopes of the height profiles decrease as indicated in Fig.~\ref{fig:height2}(a), because
the average is taken along the $y$ axis. When the average is taken for the $z$ axis as $y(z)$, 
the profile bends upwards.
A similar axis dependence is obtained for the profile of cup-shaped membrane patches~\cite{nogu19}.

\begin{figure}
\includegraphics{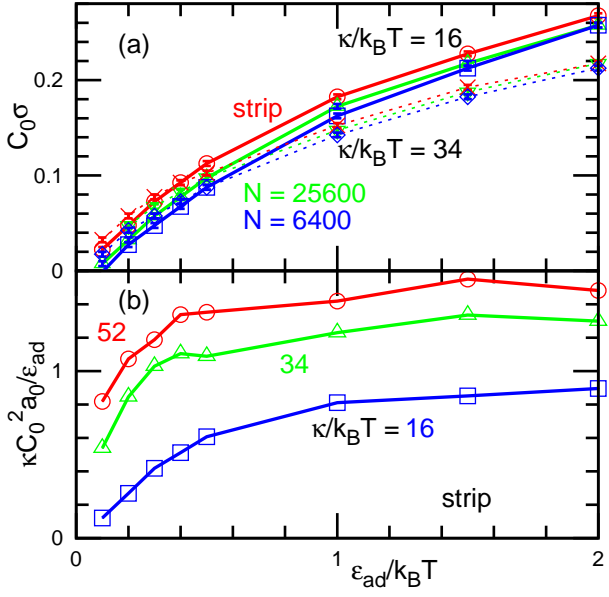}
\caption{
Dynamic phase boundary of the membrane detachment. 
Above the curves, membranes are detached from the substrate and form vesicles.
(a) The solid and dashed lines represent the data at $\kappa/k_{\rm B}T=16$, and $34$, respectively.
The red ($\circ, \times$), green ($\triangle,\triangledown$), and blue ($\square,\diamond$) colors 
indicate the data for the membrane strips and disks at $N=25~600$ and $6400$, respectively. 
(b) The detachment curvature is normalized as $\kappa C_0^2a_0/\varepsilon_{\rm ad}$
for the membrane strips at $\kappa/k_{\rm B}T=16$ ($\square$), $34$ ($\triangle$), and $52$ ($\circ$).
}
\label{fig:pde}
\end{figure}

\begin{figure}
\includegraphics{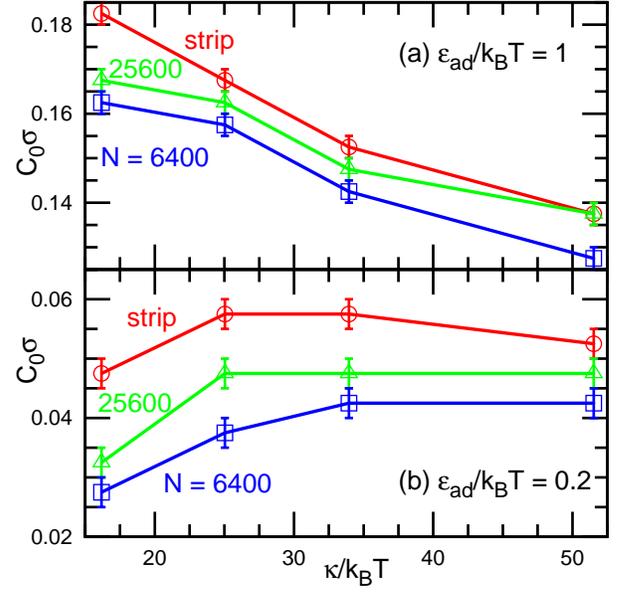}
\caption{
Bending rigidity $\kappa$ dependence of the detachment boundary at (a) $\varepsilon_{\rm ad}/k_{\rm B}T=1$
and (b) $\varepsilon_{\rm ad}/k_{\rm B}T=0.2$.
From top to bottom:  membrane strips and disks at $N=25~600$ and $6400$.
}
\label{fig:pdk}
\end{figure}

\begin{figure}
\includegraphics{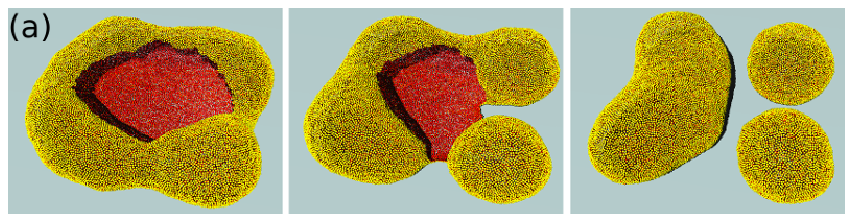}
\includegraphics{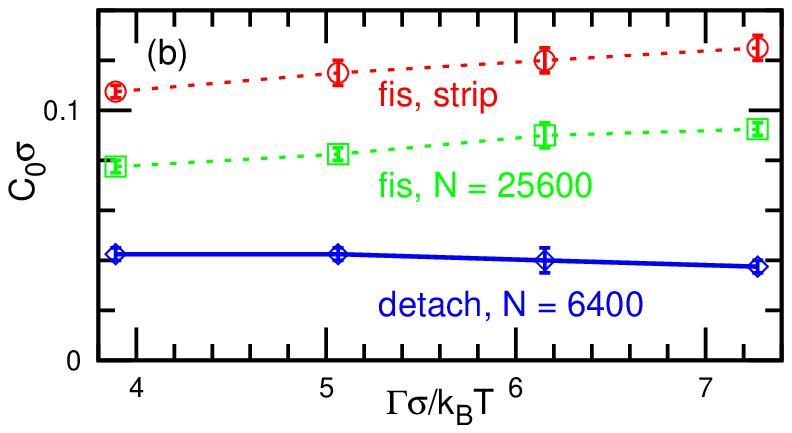}
\caption{
Edge line tension, $\Gamma$, dependence of the phase boundary of the membrane detachment and fission
at  $\kappa/k_{\rm B}T=34$.
(a) Sequential snapshots of membrane fission at $t/\tau= 14~000$, $17~4000$, and $22~000$
from a membrane disk for $C_0\sigma=0.08$, $\varepsilon_{\rm ad}/k_{\rm B}T=0.2$, and $N=25~600$.
(b) The dashed lines with ($\circ$) and ($\square$)  represent the fission boundary for 
the membrane strips at  $\varepsilon_{\rm ad}/k_{\rm B}T=0.1$ and the membrane disks at  $\varepsilon_{\rm ad}/k_{\rm B}T=0.2$ and $N=25~600$, respectively.
One or multiple vesicles are formed below or above these curves, respectively.
The solid line with ($\diamond$) represents the detachment boundary of the membrane disks
at $\varepsilon_{\rm ad}/k_{\rm B}T=0.2$ and $N=6400$.
} 
\label{fig:fis}
\end{figure}

The boundary of the membrane detachment is shown in Figs.~\ref{fig:pde}, \ref{fig:pdk}, and \ref{fig:fis}.
Above or below the boundary curves, the membranes are detached or remained on the substrate, respectively.
As expected, a higher curvature $C_0$ is required for the detachment from a stronger adhesion (a greater $\varepsilon_{\rm ad}$).
As pointed out in Ref.~\citenum{boye18}, it is attributed to the competition between the bending energy $\kappa C_0^2/2$ and 
adhesion energy $w_{\rm ad}=\varepsilon_{\rm ad}/a_0$ per unit membrane area.
To examine this relation, the boundary is normalized as $\kappa C_0^2a_0/\varepsilon_{\rm ad}$ in Fig.~\ref{fig:pde}(b).
When the adhesion is stronger than the thermal fluctuations ($\varepsilon_{\rm ad} \gtrsim k_{\rm B}T$),
the plots exhibit approximately constant values, 
so that the detachment boundary can be determined by the ratio of the bending and adhesion energies. 
However, for weaker adhesions, the detachment occurs at a smaller $C_0$, where the thermal fluctuations  are not negligible.
As the bending rigidity $\kappa$ increases, 
the detachment curvature decreases at $\varepsilon_{\rm ad}/k_{\rm B}T=1$ [see Fig.~\ref{fig:pdk}(a)].
However, interestingly, it increases for $\kappa/k_{\rm B}T \lesssim 30$ 
at $\varepsilon_{\rm ad}/k_{\rm B}T=0.2$ [see Fig.~\ref{fig:pdk}(b)].
The thermal membrane undulation is suppressed by the high bending rigidity.
Thus, under the weak-adhesion condition, the membrane is detached when the thermal undulation overcomes the adhesion.

The detachment curvature is independent of the edge tension, $\Gamma$, for the membrane strips,
because the edge length does not vary during the initial detachment process.
For a small disk with $N=6400$, the detachment curvature increases  slightly with increasing $\Gamma$ as shown 
as a solid line in
Fig.~\ref{fig:fis}(b), where the edge length decreases during the detachment.
Due to this effect,
 smaller membrane patches exhibit lower detachment curvatures [see Figs.~\ref{fig:pde}(a) and \ref{fig:pdk}].

It should be noted that the membrane fission into vesicles is slightly suppressed by an increase in $\Gamma$
as indicated by dashed lines in Fig.~\ref{fig:fis}(b). 
Above the boundary curves, the multiple vesicles are formed as shown in  Fig.~\ref{fig:fis}(a). 
Between the detachment and fission curves, the membrane patch closes into a single vesicle.
The vesicle size resulted in the fission increases with increasing $\Gamma$.
The mean vesicle size is $\langle n_{\rm ves} \rangle = 530$, $690$, and $860$
for $\Gamma\sigma/k_{\rm B}T= 3.89$, $5.1$, and $6.2$, respectively, 
under the conditions of Fig.~\ref{fig:cn}.
The high edge tension accelerates the membrane closure and suppresses an increase in the edge line length for the fission.
This is opposite to the case of a cup-to-vesicle transition, in which 
the high edge tension reduces the membrane sizes for the vesicle closure~\cite{nogu19}.

In the case that the substrate is removed at $t=0$,
the formation of one and multiple vesicles similarly occurs at low and high spontaneous curvatures, respectively.
Experimentally, vesicles can be produced by the hydration of dry lipid films~\cite{reev69,yama07}.
If one side of the membrane is fabricated to induce a spontaneous curvature by mean of polymer anchoring and so on,
the fission of a detached membrane patch can lead to the formation of small vesicles.

\begin{figure}
\includegraphics{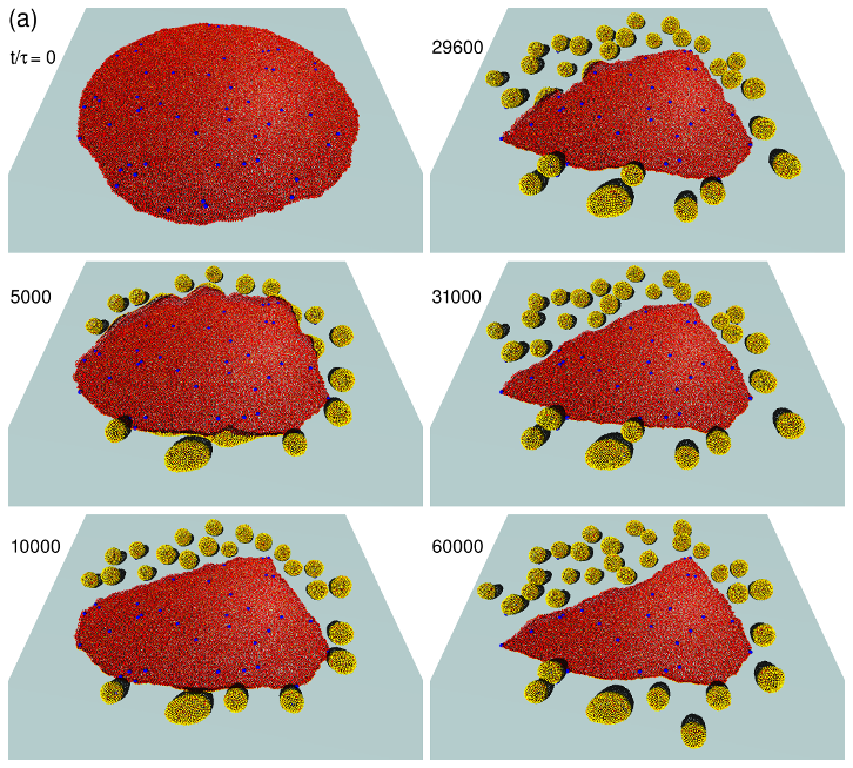}
\includegraphics{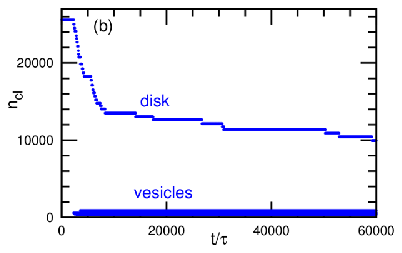}
\caption{Detachment dynamics of a pinned membrane at $\phi_{\rm pin}=0.002$,
 $C_0\sigma=0.16$, $\kappa/k_{\rm B}T=34$, $\varepsilon_{\rm ad}/k_{\rm B}T=1$, $\varepsilon_{\rm mb}/k_{\rm B}T=6$,   and $N=25~600$.
(a) Sequential snapshots at $t/\tau=0$, $5000$, $10~000$, $29~600$, $31~000$, and $60~000$.
The pinned particles are displayed as blue spheres larger than the mobile particles for clarity.
(b) Time evolution of cluster sizes $n_{\rm cl}$.
}
\label{fig:pin50}
\end{figure}

\begin{figure}
\includegraphics{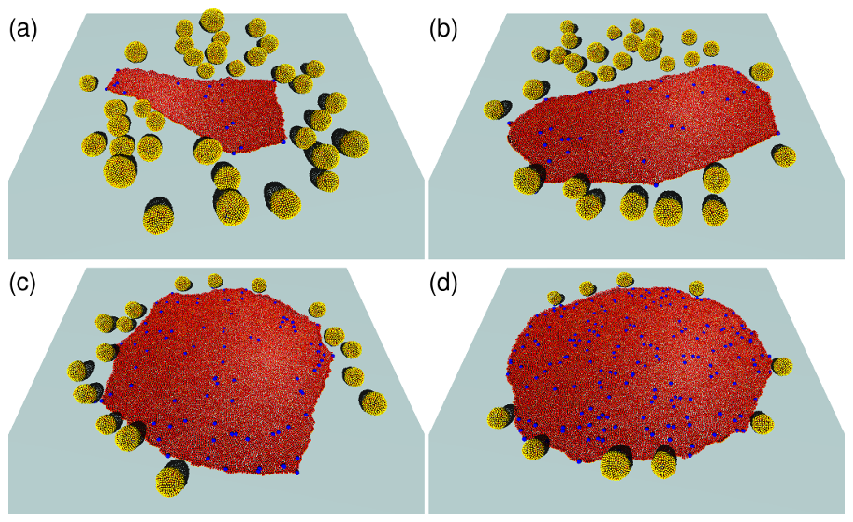}
\includegraphics{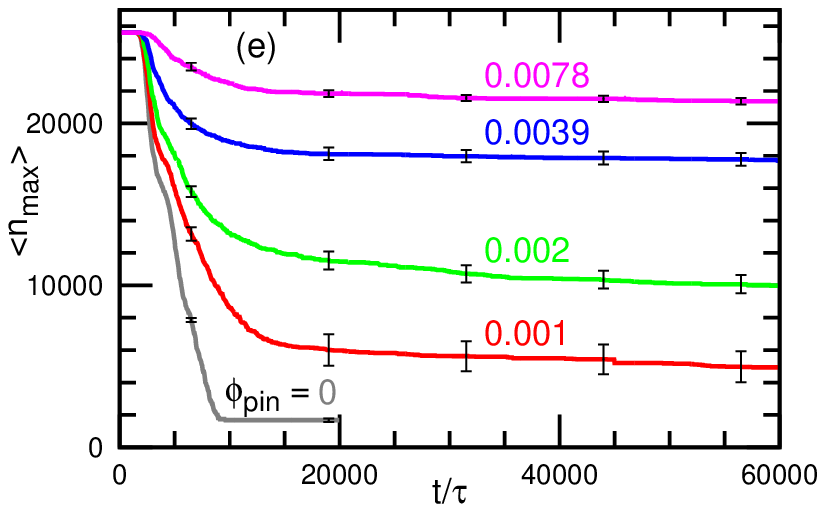}
\caption{
Dependence of the detachment dynamics on pin density $\phi_{\rm pin}$
at $C_0\sigma=0.16$, $\kappa/k_{\rm B}T=34$, $\varepsilon_{\rm ad}/k_{\rm B}T=1$, $\varepsilon_{\rm mb}/k_{\rm B}T=6$, and $N=25~600$.
(a)--(d) Snapshots at  $t/\tau=60~000$ for (a) $\phi_{\rm pin}=0.001$, 
(b) $\phi_{\rm pin}=0.002$, (c) $\phi_{\rm pin}=0.0039$, and (d) $\phi_{\rm pin}=0.0078$.
(e) Time evolution of the average of the maximum cluster size $\langle n_{\rm max}\rangle$
for  $\phi_{\rm pin}=0$, $0.001$, $0.002$, $0.0039$, and $0.0078$.
}
\label{fig:pins}
\end{figure}

\section{Detachment of Pinned Membrane} \label{sec:pin}

Next, we describe the pinning effects on the detachment.
Figures \ref{fig:pin50} and \ref{fig:pins} show the detachment dynamics of the pinned membrane patches
at $\varepsilon_{\rm ad}/k_{\rm B}T=1$.
When the membrane edge approaches a pinned particle,
it locally suppresses the detachment  [see Fig.~\ref{fig:pin50} and  Movie~S4 provided in ESI].
As the membrane edges become trapped by the pinned particles,
the patch adopts a polygonal shape and the pinned particles are located at the vertices.
Occasionally, vesicles are still formed under fluctuations of the excess membrane area [see Fig.~\ref{fig:pin50}].
The mean vesicle size becomes slightly smaller than the unpinned membranes.

As the pin density, $\phi_{\rm pin}$, increases,
larger membrane patches remain on the substrate [see Fig.~\ref{fig:pins}].
At the low densities of $\phi_{\rm pin}=0.001$ and $0.002$, the remaining membrane area varies significantly:
the entire membrane is detached  in $7$ out of $20$ runs at $\phi_{\rm pin}=0.001$.
At  $\phi_{\rm pin} \geq 0.0039$, more than half of the membrane area still adheres to the substrate.
Thus, a small amount (less than $1$\%) of pinning can significantly suppress the membrane detachment.

\begin{figure}
\includegraphics{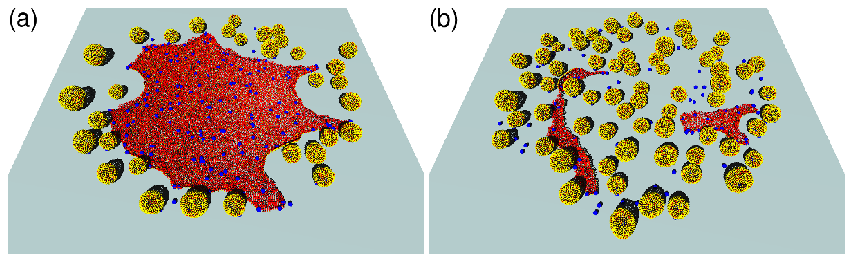}
\includegraphics{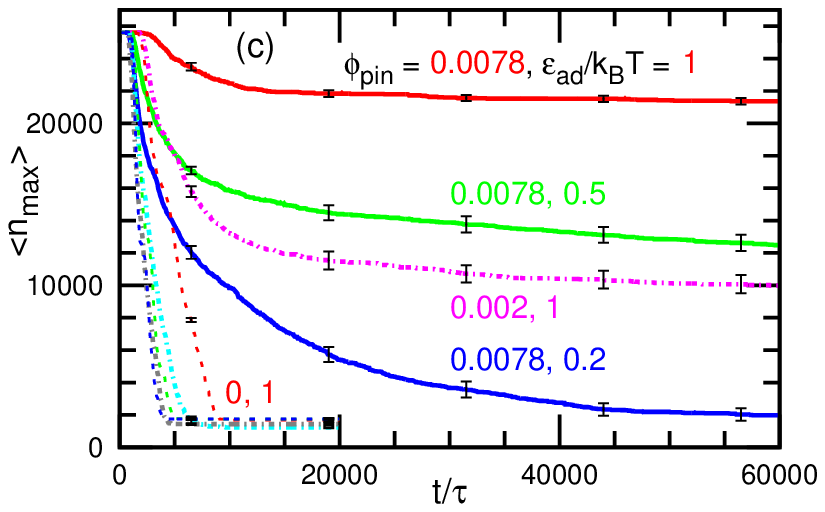}
\caption{
Dependence of the detachment dynamics on adhesion strength $\varepsilon_{\rm ad}$
at $C_0\sigma=0.16$, $\kappa/k_{\rm B}T=34$, $\varepsilon_{\rm mb}/k_{\rm B}T=6$, and $N=25~600$.
(a),(b) Snapshots for (a) $\varepsilon_{\rm ad}/k_{\rm B}T=0.5$ and (b) $\varepsilon_{\rm ad}/k_{\rm B}T=0.2$ 
at $\phi_{\rm pin}=0.0078$ and  $t/\tau=60~000$.
(c) Time evolution of the average of the maximum cluster size $\langle n_{\rm max}\rangle$.
The dotted, dashed-dotted, solid lines represent the data for $\phi_{\rm pin}=0$, $0.002$, and $0.0078$, respectively.
From top to bottom:  $\varepsilon_{\rm ad}/k_{\rm B}T=1$, $0.5$, and $0.2$.
}
\label{fig:pinad}
\end{figure}

Next, the effects of the adhesion strength $\varepsilon_{\rm ad}$ is investigated as shown in Fig.~\ref{fig:pinad}.
As $\varepsilon_{\rm ad}$ decreases, the detachment becomes faster and the whole membrane can be detached for the low density $\phi_{\rm pin}=0.002$.
Several pins are separated from the membrane, since stronger force can be exerted to a pinned particle for smaller $\varepsilon_{\rm ad}$.
Interestingly, the membrane patch can form concave edges as well as straight edges [see Fig.~\ref{fig:pinad}(a) and (b)].
Similar concave edges were observed in the experiments~\cite{boye17,boye18}. 
The edges are bent and more membrane region is detached from the substrate.
The detachment boundary is $C_0^{\rm bd}\sigma=0.15$ at $\varepsilon_{\rm ad}/k_{\rm B}T=1$.
Hence, the simulation condition,  $C_0\sigma=0.16$, is close to the boundary
and becomes more distant with decreasing  $\varepsilon_{\rm ad}$.
In such far-from-equilibrium conditions, the membrane can be detached further even from the concave edge.

\section{Summary and Discussion} \label{sec:sum}

The detachment dynamics of membrane patches from a flat substrate were simulated.
As the spontaneous curvature $C_0$ increases,
the membrane edge  bends upwards more significantly.
When $C_0$ is higher than the threshold value,
the membrane is detached from the edge and forms vesicles.
The threshold curvature is $C_0 \sim \sqrt{w_{\rm ad}/\kappa}$ for the strong-adhesion conditions
but decreases for the weak-adhesion conditions.
For small membrane patches, this curvature slightly decreases with an increase in the edge line tension.
The pinning of the membranes onto the substrate
locally suppresses the membrane detachment and straight or concave membrane edges are formed between the pinned points.

Boyes' experiments~\cite{boye17,boye18} showed several types of detachment dynamics.
The membrane blebbing induced by the  annexins A1 and A2 
is similar to the initial process of the vesicle formation.
However, the membrane pinching-off into a vesicle is prevented.
It is likely that the binding of two neighboring membranes stabilizes the neck of the bleb by means of the annexins.
Thus, we consider that the  annexins A1 and A2 engender isotropic spontaneous curvature of the membrane
and also cause the binding of two membranes.
They observed the membrane rolling for four types of the annexins (A3, A4, A5, and A13).
We obtained the rolling only for strips shorter than the wavelength of the unduloid deformation, $2\pi/C_0$.
Hence, we conclude that the isotropic spontaneous curvature cannot induce the membrane rolling, and
these annexins presumably induce the anisotropic spontaneous curvature.
Moreover, membrane solidification is also a possible cause of the rolling for the annexins A5,
since these form an ordered assembly on the membrane.
Solid membranes cannot form vesicles.
It is noted that an ordered solid membrane can exhibit  anisotropic spontaneous curvature but it is not orthogonal,
{\t e.g.}, $0$, $2\pi/3$, $4\pi/3$ can be preferred direction to bend for a hexagonal lattice.
If these proteins have the binding ability of the neighboring membranes,
the membrane binding reinforces the roll structure.
In their experiments, the concave edges are often formed.
In our simulations, concave membrane edges are obtained in the pinned membranes under the weak substrate adhesion.
Therefore, the pinning plays an essential role to form concave edges.

In this study, the Langevin dynamics is employed so that the hydrodynamic interactions are neglected.
The membrane detachment is a normal motion to the membrane while
the vesiculation is  also accompanied by the tangential motion.
When the hydrodynamic interactions are taken into account,
the tangential motion becomes relatively slower as the membrane viscosity increases compared to the solvent viscosity~\cite{naka18}.
Thus, it is expected that the vesiculation occurs later for higher membrane viscosity.

Our study has demonstrated that the membrane detachment from the substrate can lead to the formation of numerous vesicles.
The size of the obtained vesicles is several times greater 
than that of the vesicle formation via self-assembly of small membrane patches.
Moreover, the dependence on the edge tension is the opposite:
at a higher edge tension,  larger and smaller vesicles are formed for the membrane detachment and assembly, respectively.
Thus, the final vesicle size is controlled kinetically as well as thermostatically.
The resultant membrane structure can be altered by the anisotropy of the spontaneous curvature and the binding of neighboring membranes.
Our study provides basic knowledge for an improved understanding of the membrane detachment dynamics including such conditions.

\begin{acknowledgments}
We thank Olivier Pierre-Louis (Univ. Lyon 1) for stimulating discussion and
acknowledge the visiting professorship program of University of Lyon 1.
This work was supported by JSPS KAKENHI Grant Number JP17K05607.
\end{acknowledgments}

\end{document}